\documentclass[pra,twocolumn]{revtex4}%
\usepackage{graphicx}
\usepackage{dcolumn}
\usepackage{bm}
\usepackage{amsmath}
\usepackage{amsfonts}
\usepackage{amssymb}%
\begin{document}
\title{Non-full rank bound entangled states satisfying the range criterion}
\author{Somshubhro Bandyopadhyay\footnote{Present address: Department of Chemistry,
University of Toronto, 80 St. George St., Toronto, ON, M5S3H6, Canada} }
\author{Sibasish Ghosh \footnote{Present address: Department of Computer Science, The
University of York, Heslington, York, YO10 5DD, UK}}
\author{Vwani Roychowdhury}
\email{{som,vwani}@ee.ucla.edu, sibasish@cs.york.ac.uk}
\affiliation{Department of Electrical Engineering, University of California, Los Angeles,
CA 90095}

\begin{abstract}
{\small {A systematic method for generating bound entangled states in any
bipartite system, with ranks ranging from five to full rank, is presented.
These states are constructed by mixing separable states with UPB (Unextendible
Product Basis) -generated PPT bound entangled states. A subset of this class
of PPT bound entangled states, having less than full rank, is shown to satisfy
the range criterion [\textit{Phys. Lett. A} \textbf{232} (1997) 333]. }}

\end{abstract}
\maketitle

%\date{today}

\begin{center}

\end{center}


\section{Introduction}

One of the recent fundamental advances in quantum information theory
\cite{qinfo}, and in particular in the theory of quantum entanglement
\cite{qent}, is the discovery of bound entangled states \cite{Phorodecki97}:
the mixed entangled states from which no pure entanglement can be obtained by
local operations and classical communication (LOCC), whatever be the number of
copies of the state being shared. Bound entangled (BE) states have been
studied extensively in the recent past \cite{pbruss,Bennett99,HHH99}, and the
primary focus has been on obtaining succinct characterizations of bound
entangled states, on deriving appropriate tools to identify bound
entanglement, and on enumerating possible applications of bound entangled
states, if any, for quantum information processing purposes. A comprehensive
understanding of BE states, however, still remains elusive. For example, while
a few systematic procedures for constructing BE states that are positive under
partial transposition (i.e., PPT BE states) have been presented
\cite{pbruss,Bennett99,HHH99}, the relative abundance and distribution of PPT
BE states in the Hilbert space is still not completely understood.

Perhaps the main difficulty in studying bound entangled states is
related to it's identification. The problem is complicated by the
fact that most bound entangled states are positive under partial
transposition, like any separable state; the existence of BE
states that might be negative under partial transposition (NPT)
has only been conjectured \cite{nptbe}. Thus a major challenge in
identifying and characterizing BE states concerns itself with the
question of whether a given PPT state is separable or inseparable.
In general, despite recent efforts \cite{lewenstein}, there are no
succinct criteria or efficient computational tools that would
determine a separable decomposition of any given PPT state, if it
exists, or otherwise would indicate that no such decomposition is
possible. An ingenious technique to get around this hurdle is
based on studying the \textit{range} of the state under
consideration \cite{Phorodecki97}. Recall that the range of a
Hermitian operator is the space spanned by the eigenvectors
corresponding to the non-zero eigenvalues. The \emph{range
criterion} of separability (RC) can be stated as follows: If a
state $\rho_{AB}$ acting on a Hilbert space is separable, then
there exists a family of product vectors $\left\vert
\psi_{i}\right\rangle_A \otimes\left\vert \phi_{i}\right\rangle_B
$ such that (a) they span the range of $\rho_{AB}$ (b) the vectors
$\left\vert \psi_{i}\right\rangle_A \otimes\left\vert
\phi_{i}^{\ast }\right\rangle_B $ span the range of $\rho^{T_{B}}$
(where the superscript $T_B$ represents the partial transposition
operation with respect to party $B$, and * denotes the complex
conjugation in the basis where the partial transposition was
performed,  ). In particular, $\left\vert \psi_{i}\right\rangle
\otimes\left\vert \phi _{i}\right\rangle $ belongs to the range of
$\rho$. It is to be mentioned here that the separability problem
(\textit{i.e.}, to test whether any given state of a composite
system is separable) has been shown to be NP-hard \cite{gurvits}.
Recently Doherty et al. \cite{doherty} has provided a complete
family of separability criteria for detecting whether a given
state of a composite system is separable or entangled. This method
is based on the possibility of extending the state to a state of
more number of parties, satisfying some symmetry conditions. For
all separable states, such extensions are always possible, but if
a state is entangled, it will definitely lack this possibility --
the fact, which can be detected after a finitely many steps in the
hierarchy of separability criteria. And this extension method can
be cast as a semi-definite programming \cite{semidefinite}. Also
it has been shown recently by Perez-Garcia \cite{perezg} that the
cross-norm characterization of separability, a necessary and
sufficient criterion for testing separability (given by Rudolph
\cite{rudolph}), can be reduced to a linear programming problem,
for fixed chosen error.

The RC is of course a necessary condition for separability,
\emph{but if a state violates the criterion it must be entangled}.
Most systematic procedures for constructing PPT BE states,
presented so far, are based on showing that the underlying
\emph{PPT states violate the RC}. For example, the first
systematic way to construct PPT BE states was provided in Ref.
\cite{Bennett99} based on the concept of unextendible product
bases, where the BE states violate the RC in an extreme way, in
the sense that there are no product states in it's range. The
range criterion, however, cannot always be applied. If the given
state is of full rank then it trivially satisfies the range
criterion. Indeed, in Ref. \cite{vw} the authors constructed a
class of full rank PPT states in $3\otimes3$ quantum systems,
%based on St$\ddot{o}$rmer matrices,
which are entangled. In fact, all classes of bound entangled states that have
been obtained so far, either (1) violate the RC and are of less than full rank
or (2) are of full rank (only such known class is the one in $3\otimes3$,
mentioned above).

One question is imminent: \textit{Does their exist PPT BE states that do not
have full rank but nevertheless satisfy the RC\ ? Moreover, can one find a
systematic procedure to obtain PPT BE states that satisfy the RC ? }

The answers to the above questions are not immediately clear. It is also not
known whether there are full-rank PPT BE states in any $d\otimes d$ system. In
our effort to identify PPT BE states that satisfy the RC, the theory of non
decomposable positive maps and entanglement witness turns out to be extremely
useful. The witness operators to detect bound entanglement was first
introduced for UPB-generated BE states in Ref. \cite{Terhal00}, and was
developed further in Ref. \cite{Lewenstein00}.

The present work addresses the construction and identification of both
non-full rank and full-rank PPT BE states that \textit{satisfy} the RC in any
$d\otimes d$ bipartite quantum system. First, in Section~\ref{3-3-sec}, we use
the UPB states in $3\otimes3$ from Ref. \cite{DiVincenzo03} to construct a
class of PPT BE states that have rank 5 (while the system has rank 9). We
prove their inseparability \emph{from the first principles}, i.e., by showing
that the bound entangled states cannot be written as a convex combination of
the pure product states in it's support, even though the support admits an
orthogonal product basis. In Section~\ref{d-d-sec}, we generalize the results
for $3\otimes3$ and show that \emph{for any $d\otimes d$ bipartite quantum
system, there are PPT bound entangled states of rank $r$, where $d^{2}-4\leq
r\leq d^{2}$, satisfying the range criterion}. In fact, we show that a much
larger set of BE states, which includes such RC-satisfying BE states as a
subset, can be constructed as convex combinations of a UPB-generated BE state
and a separable state, which is a projector on the space spanned by a subset
of the UPB's. A proof of inseparability of these states is obtained by
constructing an appropriate entanglement witness, which allows us to
explicitly calculate well-defined ranges of the parameter used in convex
combination, such that all states in the range are PPT BE. This construction
leads to a new class of PPT BE states in any $d\otimes d$ bipartite system,
with rank ranging from 5 to $d^{2}$; only a subset of this is proven to
satisfy the RC. Note that this construction also yields \emph{full-rank PPT BE
states in any $d\otimes d$ bipartite quantum system}.

\section{Non-Full-Rank BE States in $3\otimes3$ Satisfying the RC}

\label{3-3-sec} We first show inseparability of a set of non-full rank PPT
states in $3\otimes3$ that satisfy the range criterion in a direct way. The
proof relies on the fact that the state cannot be written as a convex
combination of the pure product states in it's support \emph{even though the
support can be spanned by an orthogonal set of pure product states}, and there
are more product states than the dimension of the support. Let $\{\left\vert
\omega_{1}\right\rangle ,\left\vert \omega_{2}\right\rangle ,\left\vert
\omega_{3}\right\rangle ,\left\vert \omega_{4}\right\rangle ,\left\vert
\omega_{5}\right\rangle \}$, be the UPB in $3\otimes3$ constructed in Ref.
\cite{DiVincenzo03}:
\begin{align}
\left\vert \omega_{1}\right\rangle  &  =\left\vert 2\right\rangle \otimes
\frac{1}{\sqrt{2}}\left(  \left\vert 1\right\rangle -\left\vert 2\right\rangle
\right)  ;\left\vert \omega_{2}\right\rangle =\left\vert 0\right\rangle
\otimes\frac{1}{\sqrt{2}}\left(  \left\vert 0\right\rangle -\left\vert
1\right\rangle \right)  ,\nonumber\\
\left\vert \omega_{3}\right\rangle  &  =\frac{1}{\sqrt{2}}\left(  \left\vert
0\right\rangle -\left\vert 1\right\rangle \right)  \otimes\left\vert
2\right\rangle ;\left\vert \omega_{4}\right\rangle =\frac{1}{\sqrt{2}}\left(
\left\vert 1\right\rangle -\left\vert 2\right\rangle \right)  \otimes
\left\vert 0\right\rangle ,\nonumber\\
\left\vert \omega_{5}\right\rangle  &  =\frac{1}{\sqrt{3}}\left(  \left\vert
0\right\rangle +\left\vert 1\right\rangle +\left\vert 2\right\rangle \right)
\otimes\frac{1}{\sqrt{3}}\left(  \left\vert 0\right\rangle +\left\vert
1\right\rangle +\left\vert 2\right\rangle \right)  .\label{3-3-upb-eq}%
\end{align}
Let
\begin{equation}
\rho_{BE}=\frac{1}{4}\left(  I-\displaystyle\sum_{i=1}^{5}\left\vert
\omega_{i}\right\rangle \left\langle \omega_{i}\right\vert \right)
\end{equation}
be the associated bound entangled state. We now show that the states,
\begin{equation}
\rho_{i}(\Omega)=\Omega\left\vert \omega_{i}\right\rangle \left\langle
\omega_{i}\right\vert +\left(  1-\Omega\right)  \rho_{BE}%
\end{equation}

for any $i$ ($1\leq i\leq5$) have the following properties:

(i) They are \emph{bound entangled states} if and only if $0\leq\Omega
<\frac{1}{5}$.

(ii) \emph{They satisfy the range criterion, i.e., }the range of $\rho
_{i}(\Omega)$ is spanned by an orthogonal product basis $\left\{  \left\vert
{\psi}_{i}\right\rangle \otimes\left\vert {\phi}_{i}\right\rangle
:i=1,2,\ldots,N\right\}  $, and that of $\left(  \rho_{i}(\Omega)\right)
^{T_{B}}$ is spanned by the product basis $\left\{  \left\vert {\psi}%
_{i}\right\rangle \otimes\left\vert {\phi}_{i}^{\ast}\right\rangle
:i=1,2,\ldots,N\right\}  $, and

(iii) The range of $\rho_{i}(\Omega)$ contains \emph{more pure product states
than its dimension}: In fact, there are exactly six product pure states in the range.

Let us first start with the case $i=1$. Since,
\begin{equation}
\rho_{BE}=\frac{1}{4}\left(  I-\left\vert \omega_{1}\right\rangle \left\langle
\omega_{1}\right\vert -{\textstyle\sum\limits_{i=2}^{5}}\left\vert \omega
_{i}\right\rangle \left\langle \omega_{i}\right\vert \right)
\end{equation}
one obtains,
\begin{align}
\rho_{1}(\Omega)  & =\frac{5\Omega-1}{4}\left\vert \omega_{1}\right\rangle
\left\langle \omega_{1}\right\vert \nonumber\\
& +\frac{5(1-\Omega)}{4}\left[  \frac{1}{5}\left(  I-{\textstyle\sum
\limits_{i=2}^{5}}\left\vert \omega_{i}\right\rangle \left\langle \omega
_{i}\right\vert \right)  \right]  \label{sepa-eq}%
\end{align}

We show below that one can find \textit{five} mutually orthogonal pure product
states in the range of the rank five projector
\begin{equation}
(I-{\textstyle\sum\limits_{i=2}^{5}}\left\vert \omega_{i}\right\rangle
\left\langle \omega_{i}\right\vert )
\end{equation}
and therefore, the state
\begin{equation}
\frac{1}{5}(I-{\textstyle\sum\limits_{i=2}^{5}}\left\vert \omega
_{i}\right\rangle \left\langle \omega_{i}\right\vert )
\end{equation}
is separable. This also implies that for all $\Omega$, such that $\frac{1}%
{5}\leq\Omega\leq1$, $\rho_{1}(\Omega)$ is a convex combination of separable
states and, hence is a separable state itself.

The first part involves counting explicitly the number of {\em pure} product states in
the support of $\rho_{1}(\Omega)$ which we show that there are only six of
them.  The proof of inseparability will then follow by showing that $\rho
_{1}(\Omega)$ cannot be expressed as a convex combination of the product
states in it's support when $0\leq\Omega<\frac{1}{5}$.

Let $H_{S}$ be the subspace spanned by the UPB, let $\left\{  \left\vert
\chi_{i}\right\rangle \right\}  _{i=1}^{4}$ be a set of pairwise orthonormal
vectors spanning the orthogonal subspace $H_{S}^{\perp}$, which is the range
of the state $\rho_{BE}$. Let $A$ be the new subspace spanned by the vectors
$\left\{  \left\vert \chi_{i}\right\rangle \right\}  _{i=1}^{4}$ and
$\left\vert \omega_{1}\right\rangle $. The support of the density operators
$\rho_{1}(\Omega)$ is therefore nothing but the subspace $A$.

Any pure product state in $3\otimes3$ can be written as
\begin{equation}
\left\vert \psi\right\rangle =\left(  \alpha\left\vert 0\right\rangle
+\beta\left\vert 1\right\rangle +\gamma\left\vert 2\right\rangle \right)
\otimes\left(  \alpha^{\prime}\left\vert 0\right\rangle +\beta^{\prime
}\left\vert 1\right\rangle +\gamma^{\prime}\left\vert 2\right\rangle \right)
,
\end{equation}
where the coefficients are complex and satisfy the normalization conditions
\begin{equation}
\left\vert \alpha\right\vert ^{2}+\left\vert \beta\right\vert ^{2}+\left\vert
\gamma\right\vert ^{2}=\left\vert \alpha^{\prime}\right\vert ^{2}+\left\vert
\beta^{\prime}\right\vert ^{2}+\left\vert \gamma^{\prime}\right\vert ^{2}=1.
\end{equation}
If $\left\vert \psi\right\rangle \in A$, we must have $\left\langle
\psi|\omega_{i}\right\rangle =0$, for $i=2,...,5.$ Using the orthogonality and
normalization conditions one can show that there can be only six pure product
states in $A,$ including $\left\vert \omega_{1}\right\rangle .$ The five other
pure product states,
\begin{align}
\left\vert \eta_{1}\right\rangle  &  =\frac{1}{\sqrt{2}}\left(  \left\vert
1\right\rangle -\left\vert 2\right\rangle \right)  \otimes\left\vert
1\right\rangle \ ,\nonumber\\
\left\vert \eta_{2}\right\rangle  &  =\frac{1}{\sqrt{2}}\left(  \left\vert
1\right\rangle +\left\vert 2\right\rangle \right)  \otimes\frac{1}{\sqrt{2}%
}\left(  \left\vert 0\right\rangle -\left\vert 1\right\rangle \right)
\ ,\nonumber\\
\left\vert \eta_{3}\right\rangle  &  =\frac{1}{\sqrt{6}}\left(  2\left\vert
0\right\rangle -\left\vert 1\right\rangle -\left\vert 2\right\rangle \right)
\otimes\frac{1}{\sqrt{2}}\left(  \left\vert 0\right\rangle +\left\vert
1\right\rangle \right)  \ ,\nonumber\\
\left\vert \eta_{4}\right\rangle  &  =\frac{1}{\sqrt{3}}\left(  \left\vert
0\right\rangle +\left\vert 1\right\rangle +\left\vert 2\right\rangle \right)
\otimes\frac{1}{\sqrt{6}}\left(  \left\vert 0\right\rangle +\left\vert
1\right\rangle -2\left\vert 2\right\rangle \right)  \ ,\nonumber\\
\left\vert \eta_{5}\right\rangle  &  =\frac{1}{\sqrt{6}}\left(  \left\vert
0\right\rangle +\left\vert 1\right\rangle -2\left\vert 2\right\rangle \right)
\otimes\left\vert 2\right\rangle \ ,
\end{align}
are mutually orthogonal and form a basis in $A$. Let us write $\left\vert
{\eta}_{i}\right\rangle =\left\vert {\psi}_{i}\right\rangle \otimes\left\vert
{\phi}_{i}\right\rangle $ for $i=1,2,\ldots,5$. From Eq. (5) we see that
$\left\vert {\psi}_{i}\right\rangle \otimes\left\vert {\phi}_{i}\right\rangle
=\left\vert {\psi}_{i}\right\rangle \otimes\left\vert {\phi}_{i}^{\ast
}\right\rangle $, as for each $i=1,2,\ldots,5$, $\left\vert {\phi}%
_{i}\right\rangle $ is a real state. Since in this case ${\rho}_{1}%
(\Omega)=\left(  {\rho}_{1}(\Omega)\right)  ^{T_{B}}$, and  $\left\{
\left\vert {\psi}_{i}\right\rangle \otimes\left\vert {\phi}_{i}\right\rangle
:i=1,2,\ldots,5\right\}  $ spans the range of ${\rho}_{1}(\Omega)$, therefore
$\left\{  \left\vert {\psi}_{i}\right\rangle \otimes\left\vert {\phi}%
_{i}^{\ast}\right\rangle :i=1,2,\ldots,5\right\}  $ spans the range of
$\left(  {\rho}_{1}(\Omega)\right)  ^{T_{B}}$. Thus we see that $\rho
_{1}(\Omega)$ satisfies the range criterion for all $0\leq\Omega\leq1$.

Next, let us consider the case where $0<\Omega<\frac{1}{5}$ and let us suppose
that the state $\rho_{1}(\Omega)$ is separable. Then it must be expressed by
the convex combination of the pure product states in it's support, which
implies
\begin{align}
\rho_{1}(\Omega) &  =\Omega\left\vert \omega_{1}\right\rangle \left\langle
\omega_{1}\right\vert +\left(  1-\Omega\right)  \rho_{BE}\nonumber\\
&  ={\textstyle\sum\limits_{i=1}^{5}}\displaystyle\eta_{i}\left\vert \eta
_{i}\right\rangle \left\langle \eta_{i}\right\vert +\omega_{1}\left\vert
\omega_{1}\right\rangle \left\langle \omega_{1}\right\vert \ ,
\end{align}
where $\omega_{1},\eta_{i}\geq0,i=1,\cdots,5$. Substituting the expression for
$\rho_{BE}$ from Eq.~(\ref{sepa-eq}), and noting that
\begin{equation}
I-{\textstyle\sum\limits_{i=2}^{5}}\left\vert \omega_{i}\right\rangle
\left\langle \omega_{i}\right\vert ={\textstyle\sum\limits_{i=1}^{5}%
}\left\vert \eta_{i}\right\rangle \left\langle \eta_{i}\right\vert
\end{equation}
one obtains
\begin{align}
\frac{5\Omega-1}{4}\left\vert \omega_{1}\right\rangle \left\langle \omega
_{1}\right\vert +\frac{1-\Omega}{4}%
%TCIMACRO{\tsum }%
%BeginExpansion
{\textstyle\sum}
%EndExpansion
\left\vert \eta_{i}\right\rangle \left\langle \eta_{i}\right\vert  &
=\nonumber\\
&
%TCIMACRO{\tsum }%
%BeginExpansion
\hspace*{-2cm}
{\textstyle\sum}
%EndExpansion
\eta_{i}\left\vert \eta_{i}\right\rangle \left\langle \eta_{i}\right\vert
+\omega_{1}\left\vert \omega_{1}\right\rangle \left\langle \omega
_{1}\right\vert \ .
\end{align}

If $\Omega<\frac{1}{5}$, then we get
\[
{\textstyle\sum\limits_{i=1}^{5}}\eta_{i}^{\prime}\left\vert \eta
_{i}\right\rangle \left\langle \eta_{i}\right\vert =\beta\left\vert \omega
_{1}\right\rangle \left\langle \omega_{1}\right\vert \ ,
\]
where $\beta>0$ and at least one $\eta_{k}^{\prime}\not =0$ ($1\leq k\leq5$).
Since $\left\langle \eta_{i}|\eta_{j}\right\rangle =\delta_{ij}$ and
$\left\langle \eta_{i}|\omega_{1}\right\rangle \not =0$, for all
$i=1,\cdots,5$, we get
\[
{\textstyle\sum\limits_{i=1}^{5}}\eta_{i}^{\prime}\left\vert \eta
_{i}\right\rangle \left\langle \eta_{i}|\eta_{k}\right\rangle =\eta
_{k}^{\prime}\left\vert \eta_{k}\right\rangle =\beta(\left\langle \omega
_{1}|\eta_{k}\right\rangle )\left\vert \omega_{1}\right\rangle \ ,
\]
which is a contradiction. Thus the states $\rho_{1}(\Omega)$ are bound
entangled if and only if $0\leq\Omega<1/5$.

Above-mentioned results about ${\rho}_{1}(\Omega)$ equally hold
good for all other values of $i \in\{1, 2, \ldots, 5\}$. That it
is true for $i = 2, 3, 4$, follows from the symmetry of the four
elements $\left|{\omega}_1\right\rangle$,
$\left|{\omega}_2\right\rangle$, $\left|{\omega}_3\right\rangle$,
and $\left|{\omega}_4\right\rangle$ of the UPB of Eqn. (1) with
respect to each other. Thus, for example, in order to study the
properties of ${\rho}_{2}(\Omega)$, we need to interchange $\left|
{\omega}_{1}\right\rangle $ and $\left| {\omega}_{2}\right\rangle
$, which can be achieved (upto some unimportant global phases) by
performing the following interchange on both the systems:
$|0\rangle\leftrightarrow |2\rangle$; in order to study the
properties of ${\rho}_{3}(\Omega)$, we need to interchange $\left|
{\omega}_{1}\right\rangle $ and $\left| {\omega}_{3}\right\rangle
$, which can be achieved (upto some unimportant global phases) by
performing first the swap operation, followed by the following
interchange on the first system: $|0\rangle\leftrightarrow
|2\rangle$; in order to study the properties of
${\rho}_{4}(\Omega)$, we need to interchange $\left|
{\omega}_{1}\right\rangle $ and $\left| {\omega}_{4}\right\rangle
$, which can be achieved (upto some unimportant global phases) by
performing first the swap operation, followed by the following
interchange on the second system: $|0\rangle\leftrightarrow
|2\rangle$; Finally, in the same way, as described above, one can
show that there exist exactly six product states, namely
$|1\rangle \otimes |1\rangle$, $\frac{1}{\sqrt{2}}(|0\rangle +
|1\rangle) \otimes |2\rangle$, $|0\rangle \otimes
\frac{1}{\sqrt{2}}(|0\rangle + |1\rangle)$, $|2\rangle \otimes
\frac{1}{\sqrt{2}}(|1\rangle + |2\rangle)$,
$\frac{1}{\sqrt{2}}(|1\rangle + |2\rangle) \otimes |0\rangle$, and
$\frac{1}{\sqrt{3}}(|0\rangle + |1\rangle + |2\rangle) \otimes
\frac{1}{\sqrt{3}}(|0\rangle + |1\rangle + |2\rangle) =
\left|{\omega}_5\right\rangle$, within the range of
${\rho}_{5}(\Omega)$. Therefore, the above-mentioned analysis for
${\rho}_{1}(\Omega)$ equally holds good for all other
${\rho}_{i}(\Omega)$'s.

\section{BE states satisfying the RC in $d\otimes d$}

\label{d-d-sec} We next generalize the preceding results for the case of
$d\otimes d$. A direct proof of inseparability from the first principles,
however, seems difficult to obtain, and instead we construct an entanglement
witness to show inseparability. Let $H$ be a finite dimensional Hilbert space
of the form $H_{A}\otimes H_{B}$. For simplicity, we assume that $\dim
H_{A}=\dim H_{B}=d.$ Let $S=\left\{  \omega_{i}=\psi_{i}^{A}\otimes\varphi
_{i}^{B}\right\}  _{i=1}^{n}$ be an UPB with cardinality $\left\vert
S\right\vert =n$. Let the projector on $H_{S}:$ the subspace spanned by the
UPB, be denoted by
\begin{equation}
P_{S}={\textstyle\sum\limits_{i=1}^{n}}\left\vert \omega_{i}\right\rangle
\left\langle \omega_{i}\right\vert
\end{equation}

Then the state proportional to the projector ($P_{S}^{\perp}$, say) on
$H_{S}^{\perp}$ is given by:
\begin{equation}
\rho_{BE}=\frac{1}{D-n}\left(  I-P_{S}\right)  =\frac{P_{S}^{\perp}}{D-n}%
\end{equation}
where $D=d^{2},$. Thus $\rho_{BE}$ is bound entangled.

Let $G$ be a subset of $S$, where $1\leq\left\vert G\right\vert \leq
n=\left\vert S\right\vert $. Let $P_{G}$ be the projector onto the Hilbert
space $H_{G}$ spanned by $G.$ By following the same construction as in the
previous section, we consider PPT states of the following form:
\begin{equation}
\rho_{G}(\Omega)=\frac{\Omega}{\left\vert G\right\vert }P_{G}+\frac{1-\Omega
}{D-n}\left(  I-P_{S}\right)  .\label{BE_G}%
\end{equation}

That is, we consider a class of PPT states by mixing a subset of the UPB's
with $\rho_{BE}$, and then show that there \emph{always exists a $\mu>0$, such
that the states defined in Eq.~(\ref{BE_G}) are bound entangled for all
$0<\Omega<\mu$.} In order to show the inseparability of the states under
consideration, we consider the following witness operator that was first
stated in Ref \cite{Lewenstein00} to detect entanglement of the edge states:%
\begin{equation}
W=P_{S}-\lambda I,\label{witness1}%
\end{equation}

where $\lambda$ is chosen as the value specified in the following result:

\textbf{Lemma 1} \cite{Terhal00} \textit{Let }$S=\left\{  \omega_{i}=\psi
_{i}^{A}\otimes\varphi_{i}^{B}\right\}  _{i=1}^{n}$\textit{ be an UPB. Then }%
\begin{align}
\lambda & =\min%
%TCIMACRO{\tsum \limits_{i=1}^{n}}%
%BeginExpansion
{\textstyle\sum\limits_{i=1}^{n}}
%EndExpansion
\left\langle \phi_{A}\phi_{B}|\omega_{i}\right\rangle \left\langle \omega
_{i}|\phi_{A}\phi_{B}\right\rangle \nonumber\\
& =\min%
%TCIMACRO{\tsum \limits_{i=1}^{n}}%
%BeginExpansion
{\textstyle\sum\limits_{i=1}^{n}}
%EndExpansion
\left\vert \left\langle \phi_{A}|\psi_{i}^{A}\right\rangle \right\vert
^{2}\left\vert \left\langle \phi_{B}|\varphi_{i}^{B}\right\rangle \right\vert
^{2}%
\end{align}
\textit{ }

\textit{over all pure states }$\left\vert \phi_{A}\right\rangle \in
H_{A},\left\vert \phi_{B}\right\rangle \in H_{B}$, \textit{ exists and is
strictly larger than }$0.$

For highly symmetric UPB's, like the one given in Eq.~(\ref{3-3-upb-eq}), it
is comparatively easier to calculate the value of $\lambda$. In Ref
\cite{Terhal00}, it was also noted that a tight lower bound on $\lambda$ can
be explicitly calculated because of the high symmetry some of the UPB's. In
fact, this lower bound has been calculated in \cite{Terhal00} for the highly
symmetric Pyramid UPB of $3\otimes3$. It is now straightforward to verify that
the operator in Eq.~(\ref{witness1}) is a witness operator. First of all note
that the operator is Hermitian. Next for any product state,
\begin{equation}
\left\vert \phi_{A},\phi_{B}\right\rangle \in H,\left\langle \phi_{A},\phi
_{B}\right\vert W\left\vert \phi_{A},\phi_{B}\right\rangle \geq0
\end{equation}

where the equality is achieved by the product state for which $\left\langle
\phi_{A},\phi_{B}\right\vert P_{S}\left\vert \phi_{A},\phi_{B}\right\rangle
=\lambda$, and from lemma 1 we know such a product state exists. Therefore,
for all separable states $\sigma$, $Tr(W\sigma)\geq0.$

Now if we consider the state in Eq.~(\ref{BE_G}), then we get
\[
Tr(W\rho_{G}(\Omega))=Tr\left(  \frac{\Omega}{\left\vert G\right\vert }%
P_{G}-\lambda\rho_{G}(\Omega)\right)  =(\Omega-\lambda).
\]

Thus, $Tr(W\rho_{G}(\Omega))<0$ when $0<\Omega<\lambda$, and hence,
\emph{$\rho_{G}(\Omega)$ is inseparable for all $0<\Omega<\lambda$}. Note that
the rank of $\rho_{G}(\Omega)$ is simply $\left(  D-n\right)  +\left\vert
G\right\vert .$ Therefore, rank of this particular class of PPT BE states
ranges from $D-n+1$ to $D$ for an UPB with $n$ elements. Since $n\leq(D-4)$
and $\left\vert G\right\vert \geq1$, $5\leq rank(\rho_{G}(\Omega))\leq D$.
Unfortunately not much can be said whether the states, $\rho_{G}(\Omega)$, in
general satisfy or violate the RC. However, as we show next, a subset of these
BE states satisfy the RC in any dimension.

{{{\textbf{Definition 1}} {\textit{An UPB is said to be real (alternatively,
an UPB is said to be with real elements) if all the coefficients of each of
the elements of the UPB, with respect to the standard basis, are real.}}}}

\textbf{Theorem 1} \textit{If } $S$ \textit{ be an UPB with \emph{real
elements in}} $d\otimes d$ \textit{and} $\left\vert S\right\vert =n,$
\textit{then the bound entangled states}
\begin{equation}
\rho_{G}(\Omega)=\frac{\Omega}{\left\vert G\right\vert }P_{G}+\left(
1-\Omega\right)  \rho_{BE}%
\end{equation}

($0<\Omega<\lambda$), \textit{satisfy the range criterion for all} $G$,
\emph{such that} $\left\vert G\right\vert \geq(n-4)$.

\textbf{Proof}. Let $H_{S-G}$ be the Hilbert space spanned by the elements
remaining in the UPB $S$, after $G$ being taken out from $S$. Let
$H_{S-G}^{\bot}$ be the orthogonal complement. Since $\left\vert G\right\vert
\geq$\textit{ }$n-4,$ then $\left\vert S-G\right\vert \leq4.$ From theorem 3
of Ref. \cite{DiVincenzo03}, it is sufficient to note that $H_{S-G}^{\bot}$
can be spanned by an orthogonal set of pure product states. Since $S$ is an
UPB with only real elements therefore the projectors $P_{G}$, $P_{S}$ and
$I-P_{S}$ are invariant under partial transposition. Hence the state $\rho
_{G}(\Omega)$ is invariant under partial transposition. Now note that the
range of $\rho_{G}(\Omega)$ is nothing but the subspace $H_{S-G}^{\bot}$ that
admits an orthonormal product basis $\left\{  \left\vert {\psi}_{i}%
\right\rangle \otimes\left\vert {\phi}_{i}\right\rangle :i=1,2,\ldots
,N\right\}  $, where $N=D-(n-|G|)$. Thus we can write the projector
$I-P_{S-G}$ on $H_{S-G}^{\bot}$ as
\begin{align}
I-P_{S-G}  & ={\textstyle\sum\limits_{i=1}^{N}}\left\vert {\eta}%
_{i}\right\rangle \left\langle {\eta}_{i}\right\vert \nonumber\\
& =\frac{D-n}{1-\Omega}\left\{  {\rho}_{G}(\Omega)+\left[  \frac{1-\Omega
}{D-n}-\frac{\Omega}{|G|}\right]  P_{G}\right\}
\end{align}

where $\left\vert {\eta}_{i}\right\rangle =\left\vert {\psi}_{i}\right\rangle
\otimes\left\vert {\phi}_{i}\right\rangle $ for $i=1,2,\ldots,N$. Taking
partial transposition (with respect to the second subsystem), we have
\begin{align}
& {\textstyle\sum\limits_{i=1}^{N}}\left\vert {\psi}_{i}\right\rangle
\left\langle {\psi}_{i}\right\vert \otimes\left\vert {\phi}_{i}^{\ast
}\right\rangle \left\langle {\phi}_{i}^{\ast}\right\vert \nonumber\\
& =\frac{D-n}{1-\Omega}\left\{  {\rho}_{G}(\Omega)+\left[  \frac{1-\Omega
}{D-n}-\frac{\Omega}{|G|}\right]  P_{G}\right\}  \nonumber\\
& ={\textstyle\sum\limits_{i=1}^{N}}\left\vert {\psi}_{i}\right\rangle
\left\langle {\psi}_{i}\right\vert \otimes\left\vert {\phi}_{i}\right\rangle
\left\langle {\phi}_{i}\right\vert
\end{align}

as $S$ is a real UPB. This implies that $\left\{  \left\vert {\psi}%
_{i}\right\rangle \otimes\left\vert {\phi}_{i}^{\ast}\right\rangle
:i=1,2,\ldots,N\right\}  $ also spans $H_{S-G}^{\bot}$, and hence it also
spans the range of $\left(  {\rho}_{G}(\Omega)\right)  ^{T_{B}}\left(  ={\rho
}_{G}(\Omega)\right)  $. Therefore $\rho_{G}(\Omega)$ satisfies the range
criterion.$\square$

Thus, the class of BE states, $\rho_{G}(\Omega)$, satisfy the RC and have
ranks $(D-4)\leq rank(\rho_{G}(\Omega))\leq D$, i.e., the above construction
provides classes of bound entangled states satisfying the RC of less than full
rank, as well as, with full rank in any dimension.

Note that the condition in Theorem 1, which states that the underlying UPB
consists of real elements, is crucial because it guarantees the invariance of
the state under partial transposition. Thus, a natural question is how to
construct real UPBs for any $n\geq2d-1,$ where $2d-1$ is the lower bound on
the dimension of any UPB in $d\otimes d$. It was proved in Ref. \cite{Alon01}
that if there is an UPB with minimum dimension then it can be realized with
real elements.
%Furthermore it was shown in Ref. \cite{Wallach00}
%that the lower bound can be realized.
Unfortunately the proof is existential and not constructive. Following a
suggestion by Smolin \cite{Smolin}, here we show that for any bipartite system
we can have a real UPB with dimension $D-4.$ We first construct it in
$4\otimes4$ and as we will see the construction can be trivially generalized
to any $d\otimes d.$

Consider the real UPB in $3\otimes3$ as provided in \cite{DiVincenzo03}, and
enumerated in Eq.~(\ref{3-3-upb-eq}). Let us now add the following states:
$\left\{  \left\vert 03\right\rangle ,\left\vert 13\right\rangle ,\left\vert
23\right\rangle ,\left\vert 33\right\rangle ,\left\vert 30\right\rangle
,\left\vert 31\right\rangle ,\left\vert 32\right\rangle \right\}  $ to the
above set. Thus we have now a set $S$ of twelve pairwise orthogonal pure
product states of $4\otimes4$. It is now impossible to find out a product
state $(a|0\rangle+b|1\rangle+c|2\rangle)\otimes(a^{\prime}|0\rangle
+b^{\prime}|1\rangle+c^{\prime}|2\rangle)$ in the orthogonal subspace
$H_{S}^{\bot}$ of $S$, because $S$ contains the UPB of Eq.~(\ref{3-3-upb-eq}).
So any pure product state (if there is any), in $H_{S}^{\bot}$ must be of the
form $(a|0\rangle+b|1\rangle+c|2\rangle+d|3\rangle)\otimes(a^{\prime}%
|0\rangle+b^{\prime}|1\rangle+c^{\prime}|2\rangle+d^{\prime}|3\rangle)$, where
at least one of $d$ and $d^{\prime}$ is non-zero. But at the same time, this
later product state must have to be orthogonal to each of the product states
$|03\rangle$, $|13\rangle$, $|23\rangle$, $|33\rangle$, $|30\rangle$,
$|31\rangle$, $|32\rangle$, which, in turn, implies that both $d$ and
$d^{\prime}$ must be zero. Hence $S$ is an UPB. As one can also see, the
construction can be trivially generalized to $d\otimes d$, and after a proper
counting, the number of elements turns out to be $D-4$.

Finally, we note that it is surprising that only one witness is sufficient to
show inseparability of such a wide range of PPT BE states. Naturally we would
like to know if the witness is optimal in the sense whether it is the best
witness to detect inseparability for our class of states. For example, given a
separable state $\rho_{sep}$, we would like to know the maximum value of
$\Omega$ for which the mixed state, $\Omega{\rho_{sep}} + (1 - \Omega){\rho
}_{BE}$ remains a BE state. For the $3\otimes3$ states discussed in Section
II, we were able to exactly find the value of $\Omega$ below which the state
is a BE state, where ${\rho_{sep}}$ is a product state. Construction of a
witness that will be optimal in this sense seems to be a difficult problem.
However we show that in detecting entanglement of our class of states, the
witness in Eq.~(\ref{witness1}) is not unique. In fact there can be infinitely
many of them. Before we give an example of another witness let us prove a
helpful lemma that bounds the inner-product between a pure entangled state and
any product state.

\textbf{Lemma 2 }\textit{Let }$\left\vert \Psi\right\rangle $\textit{ be a
pure entangled state written in the Schmidt form: }%
\begin{equation}
\left\vert \Psi\right\rangle ={\textstyle\sum\limits_{i=1}^{k}}\gamma
_{j}\left\vert j\right\rangle _{A}\left\vert j\right\rangle _{B}%
\end{equation}

\textit{where the Schmidt rank} $k,2\leq k\leq d.$ \textit{ Let }$\left\vert
\gamma\right\vert ^{2}=\max\left\{  \left\vert \gamma_{j}\right\vert
^{2}\right\}  .$\textit{ Then for all normalized product states }$\left\vert
\phi_{A}\right\rangle \otimes\left\vert \phi_{B}\right\rangle ,$\textit{ }%
\begin{equation}
\left\vert \left\langle \Psi|\phi_{A}\otimes\phi_{B}\right\rangle \right\vert
^{2}\leq\left\vert \gamma\right\vert ^{2}%
\end{equation}
\textit{ }

\textbf{Proof}. We can write
\begin{align}
\left\vert \left\langle \Psi|\phi_{A}\otimes\phi_{B}\right\rangle \right\vert
^{2}  & =\left\vert \sum\gamma_{j}\left\langle \phi_{A}|j\right\rangle
\left\langle \phi_{B}|j\right\rangle \right\vert ^{2}\nonumber\\
& \leq\left\vert \gamma\right\vert ^{2}\left\vert \sum\left\langle \phi
_{A}|j\right\rangle \left\langle \phi_{B}|j\right\rangle \right\vert ^{2}%
\leq\left\vert \gamma\right\vert ^{2}%
\end{align}

and  using Schwartz inequality and the facts that $\sum\left\vert \left\langle
\phi_{A}|j\right\rangle \right\vert ^{2}\leq1$, $\left\vert \sum\left\langle
\phi_{A}|j\right\rangle \right\vert ^{2}\leq1$.$\square$

Let $\left\vert \Phi\right\rangle $ be a pure entangled state belonging to
$H_{S}^{\perp}$, where $S$ is an UPB. Let $\left\vert \gamma\right\vert ^{2}$
be the absolute square of it's largest Schmidt coefficient. Consider now the
Hermitian operator:
\begin{equation}
W=P_{S}-\frac{\lambda}{\left\vert \gamma\right\vert ^{2}}\left\vert
\Phi\right\rangle \left\langle \Phi\right\vert
\end{equation}

Then from Lemmas 1 and 2 it follows that for all product state $\left\vert
\phi_{A}\right\rangle \otimes\left\vert \phi_{B}\right\rangle \in H_{A}\otimes
H_{B},$ Tr$\left(  W\left\vert \phi_{A}\right\rangle \left\langle \phi
_{A}\right\vert \otimes\left\vert \phi_{B}\right\rangle \left\langle \phi
_{B}\right\vert \right)  \geq0.$ Consider the states defined by Eq.
\ref{BE_G}. It follows that
\begin{equation}
Tr\left(  W\rho\right)  =\frac{\Omega\left[  \left\vert \gamma\right\vert
^{2}\left(  D-n\right)  +\lambda\right]  -\lambda}{\left\vert \gamma
\right\vert ^{2}\left(  D-n\right)  }%
\end{equation}

This is negative when%

\begin{equation}
\Omega<\frac{\lambda}{\left\vert \gamma\right\vert ^{2}\left(  D-n\right)
+\lambda}%
\end{equation}

Let us note that the choice of any pure state $|\Phi\rangle$, that belongs to
$H_{S}^{\perp}$, works for our construction. However we also wish to maximize
the range over which the state is bound entangled. For example, the
above-mentioned entanglement witness $W=P_{S}-\frac{\lambda}{\left\vert
\gamma\right\vert ^{2}}\left\vert \Phi\right\rangle \left\langle
\Phi\right\vert $, will be better than the entanglement witness given in
(\ref{witness1}) (so far as detection of the bound entanglement in the state
of equation (\ref{BE_G}) is concerned), provided $|\gamma|^{2}<\frac
{1-\lambda}{D-n}$. This can be done by doing a minimization over the set of
all $\left\vert \gamma\right\vert ^{2}$ and thereby choosing the corresponding
pure state. We leave the construction of such witnesses as a future research problem.

\section{Concluding Remarks}

We have studied PPT BE states for bipartite quantum systems, and have provided
a systematic method of obtaining bound entangled states in any bipartite
system with ranks ranging from five to full rank. We have also constructed a
class of entanglement witness that detects the inseparability of our class of
PPT states. We have also shown that a subset of our class having less than
full rank satisfies the range criterion. This enabled us to provide a
qualitative classification of PPT BE states based on rank and
satisfaction/violation of range criterion. For a very specific class of states
(i.e., in $3\otimes3$) we have been able to prove the inseparability from the
first principles by showing that the bound entangled states cannot be written
as a convex combination of the product states in it's support even though the
support admits an orthogonal product basis and more product states than the
dimension of the support.

{\textbf{Acknowledgement :} Part of this work was done when S. B. was a
post-doctoral fellow at Electrical Engineering Department, UCLA, and S. G. was
visiting this department. This work was sponsored in part by the U.S. Army
Research Office/DARPA under contract/grant number DAAD
19-00-1-0172, and in part by the NSF under contract number CCF-0432296.}

\end{document}